\documentclass[aps,prl,floatfix,amsmath,amssymb,reprint]{revtex4-1}
\usepackage{graphicx}
\usepackage{color}
\usepackage{microtype}

\usepackage[colorlinks=true, allcolors=blue]{hyperref}

\def\equationautorefname#1#2\null{Eq.#1(#2\null)}

\newcommand{\Vpump}{\hat{V}_\mathrm{p}}
\newcommand{\Edetm}{\mathbf{E}_\mathrm{D}^-}
\newcommand{\Edetp}{\mathbf{E}_\mathrm{D}^+}

\begin{document}
\title{Non-Hermitian Anharmonicity Induces Single-Photon Emission}
\author{Anael Ben-Asher}
\email{anael.benasher@uam.es}
\affiliation{Departamento de  F{\'\i}sica Te{\'o}rica de la Materia Condensada and Condensed Matter Physics Center (IFIMAC), Universidad Aut{\'o}noma de Madrid, E28049 Madrid, Spain}

\author{Antonio I. Fern{\'a}ndez-Dom{\'\i}nguez}
\affiliation{Departamento de F{\'\i}sica Te{\'o}rica de la Materia Condensada and Condensed Matter Physics Center (IFIMAC), Universidad Aut{\'o}noma de Madrid, E28049 Madrid, Spain}

\author{Johannes Feist}
\affiliation{Departamento de  F{\'\i}sica Te{\'o}rica de la Materia Condensada and Condensed Matter Physics Center (IFIMAC), Universidad Aut{\'o}noma de Madrid, E28049 Madrid, Spain}

\begin{abstract}
Single-photon sources are in high demand for quantum information applications. A paradigmatic way to achieve single-photon emission is through anharmonicity in the energy levels, such that the absorption of a single photon from a coherent drive shifts the system out of resonance and prevents absorption of a second one. 
We identify a novel mechanism for single-photon emission through non-Hermitian anharmonicity, i.e., anharmonicity in the losses instead of in the energy levels. 
We demonstrate the mechanism in two types of systems,  including a feasible setup consisting of a hybrid metallodielectric cavity weakly coupled to a two-level emitter, and show that it induces high-purity single-photon emission at high repetition rates. 
\end{abstract}
\maketitle

The generation and manipulation of nonclassical light are essential for light-based quantum information technologies~\cite{o2009photonic}.
Nonclassical light is characterized by photons that are correlated with each other, and therefore, their arrival times are dependent.
The likelihood to detect coincident photons is usually quantified by the normalized high-order correlation functions at zero delay, $g_{\tau=0}^{(n\geq2)}$~\cite{loudon1973quantum}.
While $g_{\tau}^{(n)}=1$ characterizes a coherent source of classical light with a Poissonian distribution of the photon arrival times,  $g_{\tau=0}^{(n=2)}< 1$ characterizes a sub-Poissonian distribution associated with quantum light. 
When $g_{\tau=0}^{(n)} \ll 1$, the source can essentially only emit a single photon at a time.
Such single-photon devices play a vital role in secure communications~\cite{gisin2002quantum,scarani2009security}, quantum computing~\cite{knill2001scheme,kok2007linear,chang2007single,shomroni2014all}, quantum metrology~\cite{giovannetti2006quantum,giovannetti2011advances,chunnilall2014metrology} and quantum sensing~\cite{degen2017quantum}.

Single-photon emission can be generated by emitters that behave like two-level systems to a good approximation, such as atoms~\cite{darquie2005controlled}, nitrogen-vacancy (NV) centers in diamonds~\cite{aharonovich2011diamond}, and quantum dots~\cite{stachurski2022single}. While these inherently nonlinear systems offer a straightforward approach for single-photon emission, they have limitations in efficiency, emission rates, coherence time, and scalability~\cite{aharonovich2011diamond,aharonovich2016solid,lee2020integrated}. As an alternative, cavity modes, which couple efficiently to light and are highly tunable, can potentially be used to achieve bright photon emission at high rates. However, they are inherently linear due to their bosonic nature, such that single-photon emission can only be achieved through physical mechanisms inducing an effective nonlinearity~\cite{lodahl2015interfacing}. In this Letter, we describe a novel mechanism for generating efficient single-photon emission at high rates using cavity modes.

\begin{figure}[t]
\includegraphics[width=\linewidth]{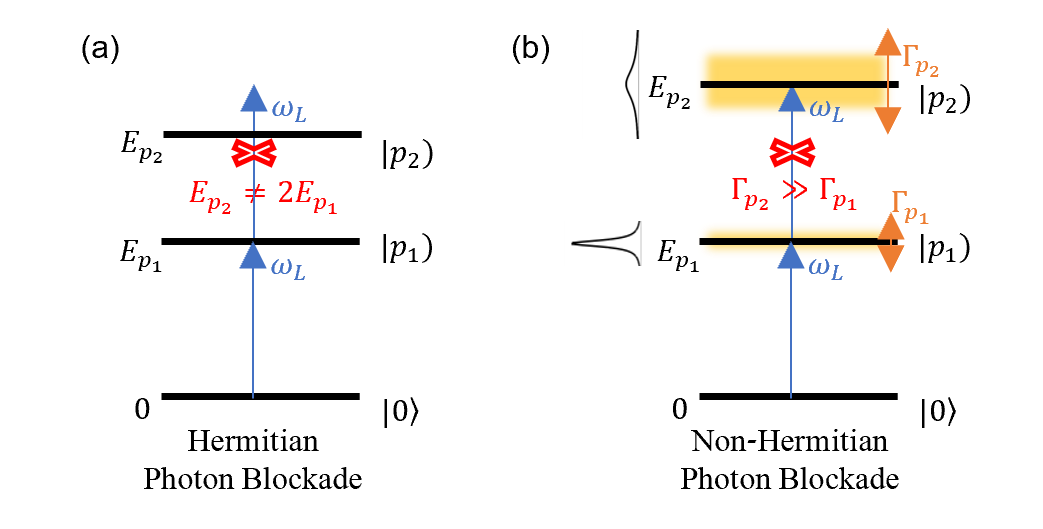}
\caption{Sketch of the two mechanisms for PB\@: (a) the  Hermitian mechanism and (b) the  NHPB mechanism  introduced in this Letter.}\label{cartoon}
\end{figure}

A prominent mechanism for the generation of single photons using cavity modes is the so-called photon blockade (PB, also known as conventional antibunching)~\cite{birnbaum2005photon,fink2008climbing,faraon2008coherent} phenomenon,  observed across various platforms, e.g., in cavity quantum electrodynamical (QED)~\cite{birnbaum2005photon,fink2008climbing} and optomechanical~\cite{rabl2011photon,stannigel2012optomechanical} systems.
This mechanism is traditionally interpreted in terms of the eigenstates of the undriven Hermitian Hamiltonian, which describes the effect of the coherent coupling of the system components but does not describe the modification of these dynamics due to losses. 
The anharmonicity of these eigenstates, i.e., the fact that the energy required to absorb one photon is different from the energy required to absorb a subsequent photon, induces the photon blockade effect.
In contrast, the novel mechanism introduced in this Letter is based on the anharmonicity of the eigenstates of the effective undriven non-Hermitian (NH) Hamiltonian arising from the Lindblad master equation for open quantum systems~\cite{visser1995solution,saez2018photon}. It becomes efficient when there is a significant difference in the losses (encoded in the imaginary parts of the NH eigenenergies) of the singly and doubly excited states of the system. Thus, we show that the effective NH Hamiltonian not only provides a mathematically convenient description including the effect of losses, but naturally points the way towards a non-Hermitian photon blockade (NHPB) mechanism.
In this aspect, it is similar to phenomena such as exceptional points and parity-time symmetry that have attracted much attention in the last decade~\cite{liertzer2012pump,regensburger2012parity,chang2014parity,sun2014experimental,doppler2016dynamically,chen2017exceptional,zhang2018phonon,zhang2018phonon,wang2020electromagnetically,wang2021coherent,ergoktas2022topological}, and also arise naturally from a NH description.

Hermitian PB is difficult to realize in lossy systems, as the anharmonicity has to be larger than the linewidths (that are determined by the loss rates) of the system eigenstates~\cite{faraon2008coherent}, which is only achieved in the strong coupling regime of cavity QED systems. In contrast, NHPB is not limited by the linewidths but instead exploits their possibly large values and operates in the weak-coupling regime of cavity QED systems, as we show below. The NHPB mechanism is thus expected to have less stringent experimental requirements than the Hermitian PB for the realization of an efficient single-photon source. Note that another antibunching mechanism, the destructive interference mechanism (also known as unconventional antibunching)~\cite{liew2010single,bamba2011origin,majumdar2012loss} also usually operates in the weak-coupling regime; however, it suffers from severe limitations for single-photon generation~\cite{zubizarreta2020conventional}.

We obtain a unified description of the PB phenomenon, which includes both the Hermitian and the NH mechanisms, using a perturbative approach to obtain a simple expression for the correlation functions under weak pumping by a continuous-wave source at frequency $\omega_L$. The derivation details and the full expressions are given in the Supplemental Material~\cite{supplemental}.
In the common situation that only one eigenstate in each excitation manifold contributes significantly to the emission, the normalized zero-delay second-order correlation function $g_{\tau=0}^{(2)}$ is given by
\begin{multline}
  \label{g2} g_{\tau=0}^{(2)}(\omega_L) \approx \left|\frac{\tilde{E}_{p_1}-\hbar\omega_L}{\frac{\tilde{E}_{p_2}}{2}-\hbar\omega_L}\right|^2 \frac{\big|(p_2|\Vpump|p_1)\big|^2}{2\big|(p_1|\Vpump|0\rangle\big|^2} \times
\\\frac{\big|(p_2|\Edetm\Edetm\Edetp\Edetp\big|p_2)|}{2\big|(p_1|\Edetm\Edetp|p_1)\big|^2}
\end{multline}
where $\Vpump$ is the pumping operator that the external driving laser couples to, transformed into  its rotating frame, $\Edetm$ is the scattered far-field operator at the detector, $|0\rangle$ is the ground state, and $|p_1)$, $|p_2)$ are the relevant  eigenstates of the NH laser-free Hamiltonian in its first- and second-excitation manifolds. Their complex eigenenergies $\tilde{E}_{j} = E_{j} - \frac{i}{2}\Gamma_{j}$, where $j$ can be either $p_1$ or $p_2$, encode both the energy position $E_{j}$ and loss rate $\Gamma_{j}$. Note that for eigenstates of a NH Hamiltonian, the notation $|\dots)$, $(\dots|$ rather than $|\dots\rangle$, $\langle\dots|$ is used to describe the right and left eigenstates~\cite{moiseyev2011non}.

The emergence of both Hermitian and non-Hermitian PB manifests itself in the first row of \autoref{g2}, which presents the ratio between the populations of $|p_2)$ and $|p_1)$. In particular, the first term of \autoref{g2} can be strongly suppressed when there is nonlinearity in the complex plane, i.e., when $\tilde{E}_{p_1}\neq\frac{\tilde{E}_{p_2}}{2}$. \autoref{cartoon} shows a sketch of these two mechanisms: While the Hermitian PB in \autoref{cartoon}(a) stems from anharmonicity in the real axis $E_{p_1}\neq\frac{E_{p_2}}{2}$, the NHPB in \autoref{cartoon}(b) occurs even when $E_{p_1}=\frac{E_{p_2}}{2}=\hbar\omega_L$ and stems from the anharmonicity in the imaginary part of the eigenenergies. 
When the decay rate of $|p_2)$ is much larger than that of $|p_1)$, i.e., when $\Gamma_{p_2}\gg\Gamma_{p_1}$, as illustrated in \autoref{cartoon}(b), the state $|p_2)$ is effectively eliminated from the dynamics such that the absorption of a second photon is prevented and $g_{\tau=0}^{(2)}(\omega_L)$ is strongly suppressed.
Note that the first term of \autoref{g2} is the ratio between two Lorentzian functions centered at $E_{p_1}$ and $E_{p_2}/2$ and whose widths are $\Gamma_{p_1}$ and $\Gamma_{p_2}/2$, respectively. These Lorentzian functions are also depicted in \autoref{cartoon}(b). They represent the energy-dependent densities of states that correspond to $|p_1)$ and $|p_2)$ in the real energy spectrum and originate from their NH character~\cite{moiseyev2011non}. This observation provides a physical explanation for the NHPB, revealing that the absorption of subsequent photons is suppressed due to a smaller density of states at twice the  laser frequency than at  the laser frequency itself.

As described above, realizing the NHPB mechanism requires a setup with a nonlinear behavior in losses.
Specifically, the narrowest  accessible  eigenstates, i.e., those with the lowest losses, in each excitation manifold of the laser-free system are the relevant $|p_1)$ and $|p_2)$ which should obey $\Gamma_{p_1}\ll\Gamma_{p_2}$ to induce it. 
This points towards a way to design single-photon sources by engineering the loss of the levels in the system, instead of designing a specific energetic structure and trying to minimize losses as the Hermitian PB suggests. 
In the following, we theoretically demonstrate this concept and present the implementation of the NHPB mechanism through the engineering of losses by tailoring the coherent coupling between different modes with a given loss within the effective NH laser-free Hamiltonian framework. 
Note that this loss-engineering approach differs from actively designing the operators in the Lindblad master equation to modify the system's interaction with the environment states~\cite{poyatos1996quantum,leghtas2015confining,lingenfelter2021unconditional,su2022nonlinear}.
We present two examples where the necessary nonlinearity is induced by different effects. The first example is a proof-of-principle model with nonlinear coupling between linear elements that facilitates analytical analysis, while the second example presents a realistic system of a hybrid metallodielectric cavity interacting with a two-level emitter (TLE) in which the nonlinearity is due to linear coupling to a nonlinear system: a TLE.

The first example is a prototypical NH Hamiltonian describing second-harmonic generation~\cite{munoz2021quantum},
\begin{equation}
   \label{Hquad}
   \hat{H}_0=\tilde{\omega}_a a^\dagger a + \tilde{\omega}_b b^\dagger b + g(a^\dagger b^2+(b^\dagger)^2 a),
\end{equation}
where $a$ ($a^\dagger$) and $b$ ($b^\dagger$) are the annihilation (creation) operators of two bosonic modes,  $\tilde{\omega}_a = 2\omega_b - \frac{i}{2} \gamma_a$ and $\tilde{\omega}_b = \omega_b - \frac{i}{2}\gamma_b$ are their complex energies, and $g$ is the quadratic coupling strength between them. 
Note that we have assumed that the real parts of the energies are exactly on two-photon resonance. 
This simple Hamiltonian can be analyzed analytically without any restriction to just a few eigenstates. 
In particular, when weakly pumping the $b$-mode ($\Vpump=b^\dagger+b$) on resonance ($\omega_L=\omega_b$), and detecting its emission ($\Edetm\propto b^\dagger$), the intensity $I$ and the normalized zero-delay second-order correlation function $g_{\tau=0}^{(2)}$ are  given by
\begin{equation}
I(\omega_L=\omega_b)\propto\frac{4}{\gamma_b^2},\quad
 g_{\tau=0}^{(2)}(\omega_L=\omega_b)=\frac{1}{(1+\eta)^2},\label{g2anal1}
 \end{equation}
where $\eta=\frac{4g^2}{\gamma_a\gamma_b}$ is the cooperativity parameter for the two modes~\cite{tanji2011interaction}.
$ g_{\tau=0}^{(2)}(\omega_L=\omega_b)$ vanishes when $\eta\gg1$. 
While large cooperativity  can be obtained either through a large $g$,  through a small $\gamma_a$, or through a small $\gamma_b$, the latter limit leads to the highest intensity and thus the most efficient single-photon emission.
This limit corresponds exactly to the NHPB mechanism, while the limits of large $g$ and small $\gamma_a$ correspond to the Hermitian PB  and the destructive interference mechanism, respectively. In the NHPB limit, the accessible first-excitation eigenstate $|p_1)$, that is the pure $b$-mode with $\Gamma_{p_1}=\gamma_b$, is narrow.
The second-excitation eigenstates, however, which arise from the interaction between the singly-excited mode $a$ and the doubly-excited mode $b$, are much broader when $\gamma_b\ll\gamma_a$. 
In particular, when the coupling is weak ($g<\frac{\gamma_a}{2\sqrt{2}}$), the decay rate of the narrowest state $|p_2)$ is given  by $\Gamma_{p_2}\approx2\gamma_b(1+\eta)$.
Thus, we obtain $g_{\tau=0}^{(2)}(\omega_L=\omega_b) \approx \big(\frac{2\Gamma_{p_1}}{\Gamma_{p_2}}\big)^2$, demonstrating the manifestation of the NHPB mechanism.
In addition, to emphasize that the low two-photon emission given by $g_{\tau=0}^{(2)}(\omega_L=\omega_b)$ indicates the low two-photon absorption characterizing the NHPB, we note that the upconversion efficiency of two absorbed photons with frequency $\omega_b$ to one emitted photon with frequency $2\omega_b$, given by $\Phi_{up}=\frac{\eta}{\eta+\gamma_a/\gamma_b}$ (see~\cite{supplemental} for  details), is very small in the NHPB limit.

The Hamiltonian in \autoref{Hquad} can, e.g., describe a mechanical oscillator (mode $b$) located exactly in the middle of a Fabry-P\'erot cavity (mode $a$), when the optical mode is strongly driven by a laser (see Refs.~\cite{xie2016single,li2022two,xie2018optically}).
Since commonly the mechanical decay rate is much smaller than the optical one~\cite{xie2016single,li2022two,xie2018optically}, the NHPB induces single-phonon emission in such a system.
For example, for $\frac{\gamma_b}{\gamma_a}=10^{-3}$ and $g=\frac{\gamma_a}{10}$, the cooperativity is $\eta=40$. Consequently, $g_{\tau=0}^{(2)}(\omega_L=\omega_b)\approx6\times10^{-3}$ (\autoref{g2anal1}), showing strong antibunching.
Note that adiabatic elimination of the mode $a$ in \autoref{Hquad} leads to a single oscillator with a nonlinear loss.
Single-phonon emission in such a system has been demonstrated in Ref.~\cite{li2020highly}.

The second example we discuss is a feasible cavity-QED-like setup involving the linear coupling of a TLE to bosonic (cavity) modes. In order to generate the anharmonicity in the complex plane that induces the NHPB, two bosonic modes with different losses are required.
Hybrid metallodielectric cavities, incorporating a narrow photonic mode and a broad plasmonic mode, provide an excellent platform that fits this requirement. These cavities have attracted much attention lately, both theoretically~\cite{yang2011hybrid,peng2017enhancing,gurlek2018manipulation,thakkar2017sculpting} and experimentally~\cite{barth2010nanoassembled,luo2015chip,cui2015hybrid}, since they combine the merits of low-loss microcavities with highly localized plasmons, thus yielding new functionalities. Very recent works~\cite{shen2022quantum,lu2021plasmonic,lu2022single,lu2022unveiling} have studied the photon statistics of a hybrid cavity coupled to a TLE and showed that it can present strong antibunching. These works related the antibunching to the destructive interference mechanism~\cite{lu2021plasmonic,lu2022single,shen2022quantum}, and/or utilized large coupling strengths or energy detuning to induce the Hermitian PB mechanism~\cite{lu2021plasmonic,lu2022single,lu2022unveiling}. In contrast, we here demonstrate that NHPB in such a system can lead to efficient single-photon emission without requiring large coupling strengths or energy detunings.

\begin{figure}[tb]
\includegraphics[width=\linewidth]{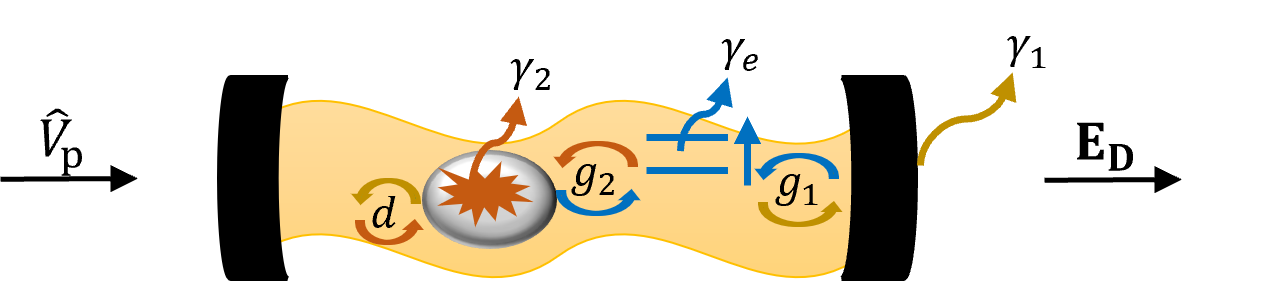}
\caption{Scheme of a hybrid metallodielectric cavity interacting with a TLE. The  parameters characterizing the system are sketched. $\Vpump$ and $\mathbf{E}_\mathrm{D}$ describe the pumping of the Fabry-P\'erot mode and the detection of its emission, respectively. 
}\label{HC}
\end{figure}

The NH Hamiltonian that describes the coupling between a TLE and two optical modes (e.g., a Fabry-P\'erot  mode and a plasmonic mode) can be written as
 \begin{eqnarray}
  \label{H}
   \hat{H}_0=(\omega_e-i\frac{\gamma_e}{2})\sigma_+\sigma_-+\sum_{n=1,2}\big[(\omega_n-i\frac{\gamma_n}{2})a^\dagger_n a_n\\\nonumber
   +g_n(\sigma_+a_n+a^\dagger_n\sigma_-)\big]+d(a^\dagger_1a_2+a^\dagger_2a_1).
 \end{eqnarray}
 Here, $\sigma_-$ ($\sigma_+$) and $a_n$ ($a_n^\dagger$) are the annihilation (creation) operators of the TLE and the two optical modes, respectively;
 $\omega_e,\omega_n$ and $\gamma_e,\gamma_n$ are respectively their energies and decay rates; $g_1,g_2$ are the coupling strengths between the TLE and each optical mode, and $d$ is the coupling strength between the two optical modes originating from the interaction between the Fabry-Pérot electric field and the plasmonic dipole moment. Note that due to the non-Hermicity of the system and the fact that the Hermitian and anti-Hermitian parts of the effective Hamiltonian do not commute, diagonalizing the Hermitian part to decouple the two optical modes would yield dissipative interactions. Consequently, there is no equivalent system of uncoupled bosonic modes that provides the same dynamics~\cite{franke2019quantization,medina2021few}. In the following, we show the importance of the coupling $d$ in inducing the NHPB mechanism.
 Without loss of generality, we assign index $1$ to the narrow (Fabry-P\'erot) mode and index $2$ to the broad (plasmonic) mode, such that $\gamma_2\gg\gamma_1$. In addition, we consider $\gamma_e\ll\gamma_{1,2}$, which describes a good emitter at low temperatures (e.g., see Ref.~\cite{wang2019turning}). \autoref{HC} depicts an illustration of the system.

The  NHPB mechanism is induced in the system above by the decoupling of the Fabry-P\'erot mode and emitter from the  plasmonic mode occurring when $g_2,d \gg g_1,\gamma_e,\gamma_1$ \cite{supplemental}. This decoupling is only present in the first-excitation manifold, forming a very narrow $|p_1)$.
The state $|p_1)$ resembles the dark state formed when several emitters interact with a cavity~\cite{del2015quantum}.
However, $|p_1)$  is optically accessible by pumping the Fabry-P\'erot mode and detecting its emission, such that $\Vpump = a_1^\dagger+a_1$ and $\Edetm\propto a_1^\dagger$. These operators describe a system configuration in which its driving from and leakage into free space is mediated by the mirrors of the Fabry-P\'erot cavity, as is presented in \autoref{HC}.
Ref.~\cite{lu2022unveiling} has discussed the formation of a very narrow $|p_1)$ in a hybrid cavity interacting with a TLE and presented its role in enhancing the efficiency of the single-photon emission occurring due to the Hermitian PB mechanism. We here show that by engineering the decay rate of the narrowest state in the second-excitation manifold, $|p_2)$, the very narrow  $|p_1)$ can be utilized to induce the NHPB mechanism.

\begin{figure}[tb]
\includegraphics[width=\linewidth]{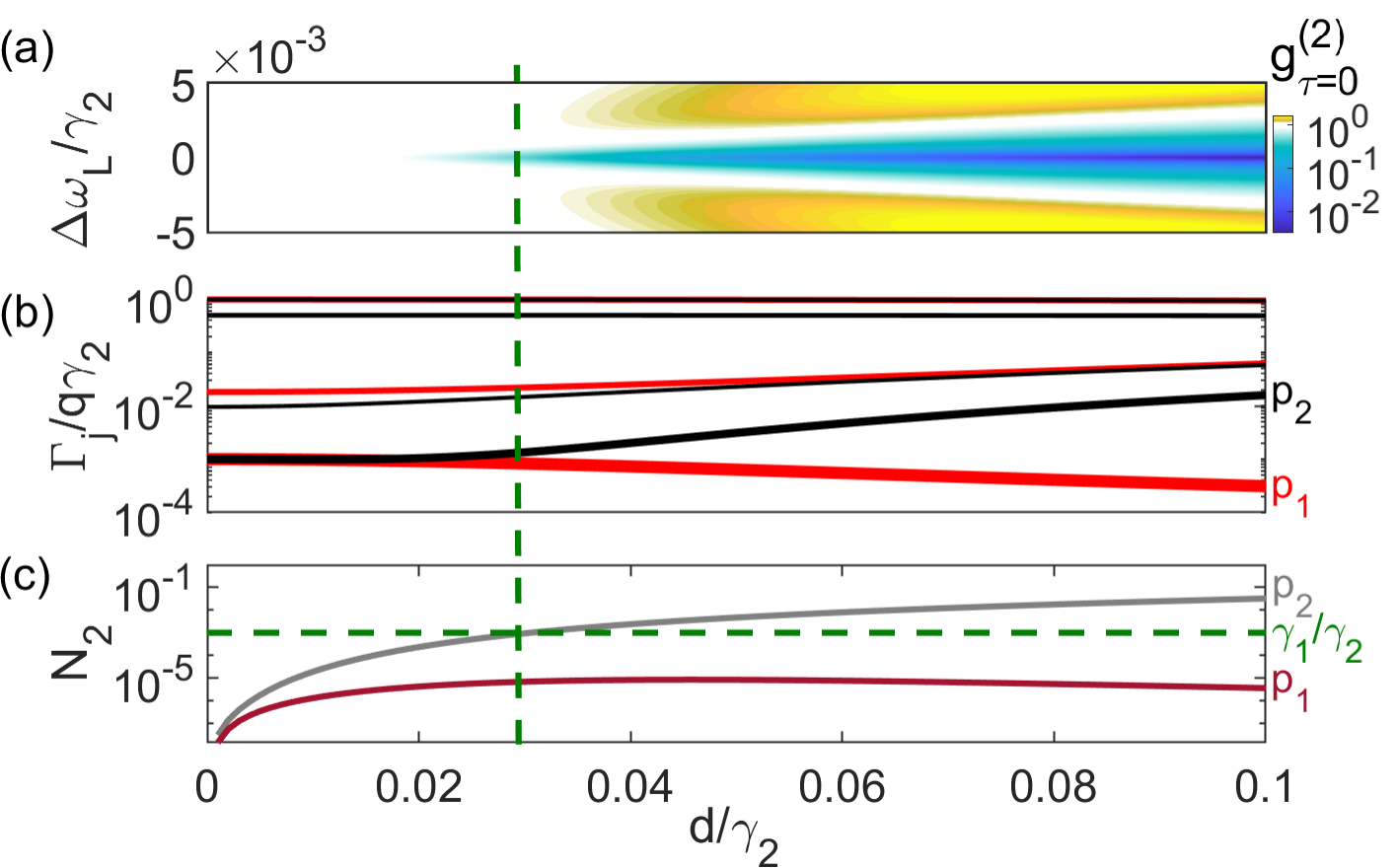}
\caption{(a)  $g_{\tau=0}^{(2)}$  as a function of   $\Delta\omega_L$, and  $d$. (b) $\Gamma_{j}/q$ of the first- (red, $q=1$) and the second- (black, $q=2$) excitation eigenstates and (c)  $|(p_1| a_2^\dagger a_2|p_1)|$ (dark red) and $|(p_2| a_2^\dagger a_2|p_2)|$ (grey) as a function of $d$.
}\label{coeff}
\end{figure}

The decay rate of $|p_2)$ is engineered by controlling its plasmonic component.
The plasmonic mode participates in $|p_2)$ through its coupling $d$ to the Fabry-P\'erot mode.
\autoref{coeff} studies the single-photon emission due to the NHPB mechanism with respect to $d$  when $\gamma_1= 10^{-3} \gamma_2$, $\gamma_e = 10^{-5}\gamma_2$, $g_1=0$, $g_2=\frac{\gamma_2}{15}$ and $\omega_e=\omega_1=\omega_2$. 
In \autoref{coeff}(a), $g_{\tau=0}^{(2)}$ is shown as a function of the laser detuning $\Delta\omega_L=\omega_e-\omega_L$ and the coupling $d$,  when pumping the Fabry-P\'erot mode and detecting its emission. 
Note that, although not shown, the intensity reaches its maximum for $\Delta\omega_L=0$ because the  coupling strengths are sufficiently small not to induce Rabi splitting. 
The observed antibunching  stems from the NHPB mechanism, i.e., from the anharmonicity in the imaginary part of the eigenenergies. 
This anharmonicity is presented in \autoref{coeff}(b), which depicts the decay rates $\Gamma_{j}$ of  the first- (red) and second-  (black) excitation eigenstates, normalized by the excitation number $q=1,2$, as a function of $d$. The decay rates of  $|p_1)$ and $|p_2)$   are in thicker lines.
\autoref{coeff}(c) plots the plasmonic components of $|p_1)$ (dark red line) and of $|p_2)$ (grey line), given by
$|(p_1| a_2^\dagger a_2|p_1)|$  and  $|(p_2| a_2^\dagger a_2|p_2)|$, respectively, as a function of $d$.
While the component of the plasmon in $|p_1)$ is kept very small, consistent with its decoupling from $|p_1)$,  its component in $|p_2)$  increases with larger $d$.
When it reaches the ratio $\frac{\gamma_1}{\gamma_2}$ (dashed horizontal line in \autoref{coeff}(c)), the decay rate of $|p_2)$ becomes significantly larger than that of $|p_1)$ (\autoref{coeff}(b)), inducing the emission of single photons (\autoref{coeff}(a)).

Similarly to the previous example,  $g_{\tau=0}^{(2)}$ can be expressed as a function of the cooperativity between the two optical modes $\eta=\frac{4d^2}{\gamma_1\gamma_2}$:
  \begin{eqnarray}
   \label{g2anal2}
   g_{\tau=0}^{(2)}(\omega_L=\omega_e)\approx\frac{1}{\eta^2}\bigg[\frac{g_2^2}{d^2}+2+4\bigg(\frac{g_2}{\gamma_2}\bigg)^2\bigg(\frac{g_2^2}{d^2}-1\bigg)\bigg]^2,
 \end{eqnarray}
 showing that large $\eta$ is required to achieve antibunching. 
\autoref{g2anal2} is the second-order approximation for the analytical expression of  $g_{\tau=0}^{(2)}(\omega_L=\omega_e)$ when $\frac{\gamma_1}{\gamma_2}\ll 1$ and $\gamma_e=0$. 
Further investigation of the dependence of the results and their robustness with respect to the system parameters is presented in the Supplemental Material~\cite{supplemental}.

\begin{figure}[tb]
\includegraphics[width=\linewidth]{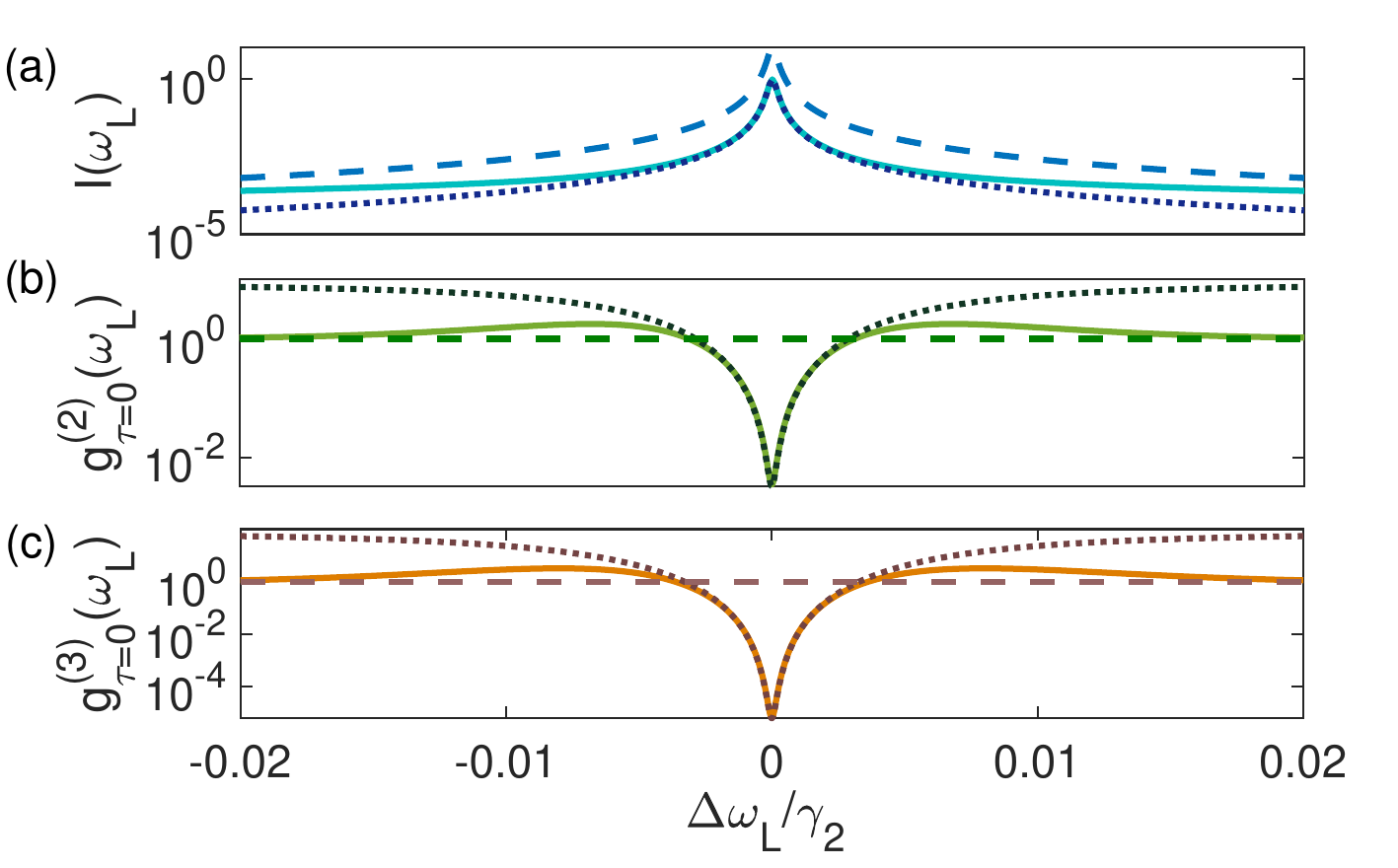}
\caption{(a)  $I$,   (b)  $g_{\tau=0}^{(2)}$ and (c) $g_{\tau=0}^{(3)}$  as a function of   $\Delta\omega_L$.
The solid lines correspond to the full calculations~\cite{supplemental}, the dotted lines consider only the narrowest eigenstate in each excitation manifold, and the dashed lines were obtained when imposing $\Gamma_{j}=q\Gamma_{p_1}$ where $q$ is the excitation number.
}\label{numericalexample}
\end{figure}

Finally, in \autoref{numericalexample}, we analyze the validity of the approximations inherent to our analysis above. 
\autoref{numericalexample}(a) and \autoref{numericalexample}(b) plot $I$ and  $g_{\tau=0}^{(2)}$, respectively, as a function of $\Delta\omega_L$ when $d=\frac{\gamma_2}{10}$ (and the other parameters are as in \autoref{coeff}). 
In addition, since single-photon emission requires the suppression of all the high-order correlations $g_{\tau=0}^{(n)}$, we present in \autoref{numericalexample}(c) the normalized zero-delay third-order correlation function $g_{\tau=0}^{(3)}$ as a function of  $\Delta\omega_L$. 
The strong dip in \autoref{numericalexample}(c) (solid line) demonstrates that the NHPB operates beyond second-order processes and suppresses multi-photon events. Moreover, the results support that the predicted antibunching is related to the generalized PB phenomenon and not  to the destructive interference effect~\cite{zubizarreta2020conventional}.
To verify that the predicted antibunching can be well-understood in terms of just two system eigenstates, $|p_1)$ and $|p_2)$, as \autoref{g2} does, we compare $I$, $g_{\tau=0}^{(2)}$ and $g_{\tau=0}^{(3)}$ obtained by including all system eigenstates (solid lines) with the values obtained  only by the narrowest eigenstates in each manifold, $|p_1)$, $|p_2)$ and $|p_3)$ (dotted lines).
As can be seen, almost perfect agreement is achieved between the two cases in the vicinity of $\Delta\omega_L=0$.
Furthermore, we show that artificially setting all the eigenstates to have the same decay rate $\Gamma_{j}=q\Gamma_{p_1}$ (dashed lines) suppresses the antibunching completely. $q$ is the excitation number, and $\Gamma_{p_1}$ is the decay rate of $|p_1)$.
This suppression demonstrates that indeed the anharmonicity in the imaginary part of the eigenenergies, i.e., the NHPB mechanism, is the origin of the observed antibunching.

To conclude, we have theoretically proposed a novel mechanism for generating high-purity single-photon emission at high repetition rates. This mechanism stems from the difference in linewidth between the absorption of one photon and  of two photons, and can be explained by the anharmonicity in the complex NH eigenenergy spectrum. 
Thus, the NHPB mechanism reveals an interesting interplay between non-Hermiticity of the system and quantum nonlinearity of the emitted light.
Importantly, it provides a reliable route to  single-photon emission even in the weak-coupling regime where couplings do not overcome  losses.
We have demonstrated the NHPB in two types of systems, including a feasible cavity-QED-like setup with realistic parameters, and showed that it indeed induces strong antibunching. 
Our findings  open the door to the realization of new single-photon devices with potential applications in quantum information,  communication and metrology. 

\begin{acknowledgments}
This work has been funded by the Spanish Ministry of Science, Innovation and Universities-Agencia Estatal de Investigación through grants PID2021-126964OB-I00, PID2021-125894NB-I00, and CEX2018-000805-M (through the María de Maeztu program for Units of Excellence in R\&D). We also acknowledge financial support from the Proyecto Sin\'ergico CAM 2020 Y2020/TCS-6545 (NanoQuCo-CM) of the Community of Madrid,  from the European Research Council through grant ERC-2016-StG-714870 and from the European Union's Horizon Europe Research and Innovation Programme through agreement 101070700 (MIRAQLS)  and No. 101034324 (CIVIS3i).
\end{acknowledgments}

\end{document}


\title{Supplemental material for: Non-Hermitian Anharmonicity Induces Single-Photon Emission}
\author{Anael Ben-Asher}
\affiliation{Departamento de Física Teórica de la Materia Condensada and Condensed Matter Physics Center (IFIMAC), Universidad Autónoma de Madrid, E28049 Madrid, Spain}

\author{Antonio I. Fernández-Domínguez}
\affiliation{Departamento de Física Teórica de la Materia Condensada and Condensed Matter Physics Center (IFIMAC), Universidad Autónoma de Madrid, E28049 Madrid, Spain}

\author{Johannes Feist}
\affiliation{Departamento de Física Teórica de la Materia Condensada and Condensed Matter Physics Center (IFIMAC), Universidad Autónoma de Madrid, E28049 Madrid, Spain}

\maketitle

 \section{Description of the correlation functions using the system eigenstates and eigenenergies}

The effective Hamiltonian in the rotating frame of a system pumped by the laser with pumping strength $\Omega$ and by the operator $\Vpump$ can be written as~\cite{saez2018photon}
\begin{equation}
\hat{H}_\mathrm{eff}=\Delta\hat{H}_0+\Omega\Vpump.
\end{equation}
Here, $\Delta\hat{H}_0=\hat{H}_0-\hbar\omega_L\hat{N}$ where $\hat{H}_0$ is the NH Hamiltonian of the laser-free system, $\hat{N}$ is a number operator that counts how many laser photons have been absorbed by the system, and $\omega_L$ is the laser frequency. The system's steady state $|\Psi_{ss}\rangle$ can be determined perturbatively in the low pumping regime ($\Omega\to0$). Expanding the solution as a power series in $\Omega$, $|\Psi_{ss}\rangle=\sum_{n=0}\Omega^n|\Psi_n\rangle$, and solving $\hat{H}_\mathrm{eff}|\Psi_{ss}\rangle=0$~\cite{saez2018photon} yields the following Born series:
\begin{equation}
|\Psi_{ss}\rangle=\sum_{n=0}\Omega^n\bigg(\frac{1}{\Delta\hat{H}_0}\Vpump\bigg)^n|0\rangle \label{ss}
\end{equation}
where $|0\rangle$ is the system's ground state. By using the spectral representation of $\Delta\hat{H}_0$, such that $\frac{1}{\Delta\hat{H}_0}=\sum_q\sum_{r_q=1}^{N_q}\frac{|r_q)(r_q|}{E_{r_q}-q\hbar\omega_L}$, we reveal the contribution of the system's eigenstates $|r_q)$ corresponding to the eigenenergies $E_{r_q}$. Here, $q$ is the excitation number (the eigenvalue of $\hat{N}$) and $N_q$ is the number of eigenstates in the $q$-excitation manifold. The c-product $(\dots|\dots)$ is used due to the system's non-Hermiticity~\cite{moiseyev2011non}. The intensity and second-order correlation function are evaluated, respectively, using the first- and second- order Born series:
\begin{eqnarray}
\label{IG2} \\ I(\omega_L)=    |\Omega|^2\bigg|\sum_{i=1}^{N_1}\sum_{j=1}^{N_1}\frac{\langle0|\Vpump|i)}{E_{i}-\hbar\omega_L}(i|\Edetm\Edetp|j)\frac{(j|\Vpump|0\rangle}{E_{j}-\hbar\omega_L}\bigg| \nonumber
   \\   \nonumber G_{\tau=0}^{(2)}(\omega_L)=|\Omega|^4\bigg|\sum_{m=1}^{N_2}\sum_{n=1}^{N_2}\sum_{i=1}^{N_1}\sum_{j=1}^{N_1}\frac{\langle0|\Vpump|i)}{E_{i}-\hbar\omega_L}\frac{(i|\Vpump|m)}{E_m-2\hbar\omega_L}  
 (m|\Edetm\Edetm\Edetp\Edetp|n)\frac{(n|\Vpump|j)}{E_{n}-2\hbar\omega_L}\frac{(j|\Vpump|0\rangle}{E_{j}-\hbar\omega_L}\bigg|
\end{eqnarray}
where $\Edetm$ is the scattered far-field operator at the detector. The normalized second-order correlation is given by $ g_{\tau=0}^{(2)}(\omega_L)=\frac{ G_{\tau=0}^{(2)}(\omega_L)}{[I(\omega_L)]^2}$~\cite{loudon1973quantum}.
Eq.~(1) in the main text is obtained when considering only one eigenstate in each excitation manifold, i.e., considering only the terms $j=i=p_1$, $n=m=p_2$ in \autoref{IG2}. $ g_{\tau=0}^{(3)}(\omega_L)$, depicted in Fig.~4 (c) in the main text, is given by $\frac{ G_{\tau=0}^{(3)}(\omega_L)}{[I(\omega_L)]^3}$ where $G_{\tau=0}^{(3)}(\omega_L)=\langle\Psi_{ss}|\Edetm\Edetm\Edetm\Edetp\Edetp\Edetp|\Psi_{ss}\rangle$ is evaluated in a similar manner as $G_{\tau=0}^{(2)}(\omega_L)$ and $I(\omega_L)$.

\section*{The upconversion efficiency in the second Harmonic generation Hamiltonian}
Considering the situation where two photons were  absorbed simultaneously by the system described by the Hamiltonian in Eq.~(2) in the main text and that is weakly pumped through  the $b$-mode, their upconversion efficiency is given by:
\begin{eqnarray}
\Phi_{up}=\frac{\langle\Psi_{ss}|a^\dagger a|\Psi_{ss}\rangle}{\langle\Psi_{ss}|a^\dagger a|\Psi_{ss}\rangle+\langle\Psi_{ss}|b^\dagger b^\dagger b b|\Psi_{ss}\rangle}.
\end{eqnarray}
 $|\Psi_{ss}\rangle$ is the  steady state  given in \autoref{ss} for $\Vpump=b^\dagger+b$,  $\omega_L=\omega_b$ and the $\hat{H}_0$ given in Eq.~(2) in the main text, $\langle\Psi_{ss}|a^\dagger a|\Psi_{ss}\rangle$ quantifies the number of photons emitted from the $a$-mode, and $\langle\Psi_{ss}|b^\dagger b^\dagger b b|\Psi_{ss}\rangle$ quantifies the amount of two photons that  were emitted simultaneously from the $b$-mode and were not upconverted. 
 The  expression given in the main text, $\Phi_{up}=\frac{\eta}{\eta+\gamma_a/
\gamma_b}$, is analytically derived using the second-order Born series for $|\Psi_{ss}\rangle$.

\section*{Non-Hermitian photon blockade (NHPB) in a hybrid cavity interacting with a two-level emitter (TLE)}

\subsection{Forming a narrow first-excitation eigenstate and a broad second-excitation eigenstate}

The Hamiltonian in Eq.~(4) in the main text is represented (when $\omega_e=\omega_1=\omega_2$) in the first-excitation manifold by the matrix:
\begin{equation}
\underline{\underline{H}}^{(1)}=
\begin{pmatrix}
\omega_e-i\frac{\gamma_e}{2} & g_{1} & g_2 \\
g_{1} & \omega_e-i\frac{\gamma_1}{2} & d \\
g_2 & d & \omega_e-i\frac{\gamma_2}{2}
\end{pmatrix}
\end{equation}
and in the second-excitation manifold by the matrix:
\begin{equation}
\underline{\underline{H}}^{(2)} =
\begin{pmatrix}
2\omega_e-i\frac{\gamma_e+\gamma_1}{2} & d &  \sqrt{2}g_{1} & g_2 & 0\\
d & 2\omega_e-i\frac{\gamma_e+\gamma_2}{2} & 0&  g_{1} & \sqrt{2}g_2 \\
\sqrt{2}g_{1} & 0&  2\omega_e-i\gamma_1 & \sqrt{2}d & 0 \\
g_2 & g_1 & \sqrt{2}d & 2\omega_e-i\frac{\gamma_1+\gamma_2}{2} & \sqrt{2}d \\
0& \sqrt{2}g_2 & 0& \sqrt{2}d & 2\omega_e-i\gamma_2
\end{pmatrix}.
\end{equation}
The basis sets used to build $\underline{\underline{H}}^{(1)}$ and $\underline{\underline{H}}^{(2)}$ are $\{\sigma_+|0\rangle, a_1^\dagger|0\rangle,a_2^\dagger|0\rangle\}$ and $\{\sigma_+a_1^\dagger|0\rangle,\sigma_+a_2^\dagger|0\rangle, \frac{(a_1^\dagger)^2}{\sqrt{2}} |0\rangle,a_1^\dagger a_2^\dagger|0\rangle, \frac{(a_2^\dagger)^2}{\sqrt{2}} |0\rangle\}$, respectively. 
We study the photon statistics of the emission from the Fabry-Pérot mode, $\Edetm \propto a_1$.

A very narrow first-excitation state $|p_1)$ can be formed, even when $d$ and $g_2$ are not negligible, while approximating  $\gamma_e\approx\gamma_1\approx0$ in two edge cases:
(i) when $g_1\to0$ and (ii) when $g_2\approx d$. In these cases,  $|p_1)$ corresponds to the  zeroth-order eigenvector:  $\Vec{v}_{p_1}^{(0)}=\tilde{N}\begin{pmatrix}
-d & g2 &0
\end{pmatrix}^t$, such that the singly-excited state of the plasmonic mode $a_2^\dagger|0\rangle$ is decoupled. $\tilde{N}$ is the normalization factor of $\Vec{v}_{p_1}^{(0)}$.
The correction to this eigenvector to first order in $g_1$, $\gamma_1$, and $\gamma_e$ is
\begin{align}
\Vec{v}_{p_1}^{(1)}=\sum_\pm\frac{16(g_1(d^2-g_2^2)+idg_2\frac{\gamma_1-\gamma_e}{2})}{A_\pm \sqrt{d^2+g_2^2}\sqrt{A_\pm^2+16(d^2+g_2^2)}}\Vec{v}^{(0)}_\pm
\end{align}
where $A_\pm=-i\gamma_2\pm\sqrt{16(d^2+g_2^2)-\gamma_2^2}$ and $\Vec{v}^{(0)}_\pm$ are the other zeroth-order eigenvectors of $\underline{\underline{H}}^{(1)}$ when $\gamma_e\approx\gamma_1\approx0$ and $g_1\to0$.
The correction $\Vec{v}_{p_1}^{(1)}$ is negligible when the following conditions are fulfilled:
\begin{align}
\frac{|\gamma_1-\gamma_e|}{4} &\ll \frac{2(g_2^2+d^2)}{\gamma_2}\label{eq:NHPB_conds}\\\nonumber
g_1|d^2-g_2^2| &\ll \frac{2(g_2^2+d^2)^2}{\gamma_2}.
\end{align}
This gives rise to the formation of a very narrow state $|p_1)$. Note that similar to Eq.~(5) in the main text, the first condition requires large cooperativity for the two optical modes $\eta=\frac{4d^2}{\gamma_1\gamma_2}$.
These conditions were derived by assuming that $\sqrt{d^2+g_2^2}\ll\frac{\gamma_2}{4}$ (weak-coupling regime) and $A_+\approx-\frac{8i(d^2+g_2^2)}{k2}$.

Forming a broad second-excitation eigenstate whose emission can be detected requires the coupling of the doubly excited state $\frac{(a_1^\dagger)^2}{\sqrt{2}} |0\rangle$ with a mode that involves excitation of the plasmonic mode. This is achieved through the coupling $d$ between the two optical modes. Fig.~3 in the main text demonstrates that with increasing $d$, the narrowest second-excitation state $|p_2)$ becomes broad due to the increasing component of the plasmonic mode in it.

\subsection{Robustness with respect to $g_2$}

\begin{figure}[ht]
\centering
\includegraphics[width=0.5\textwidth]{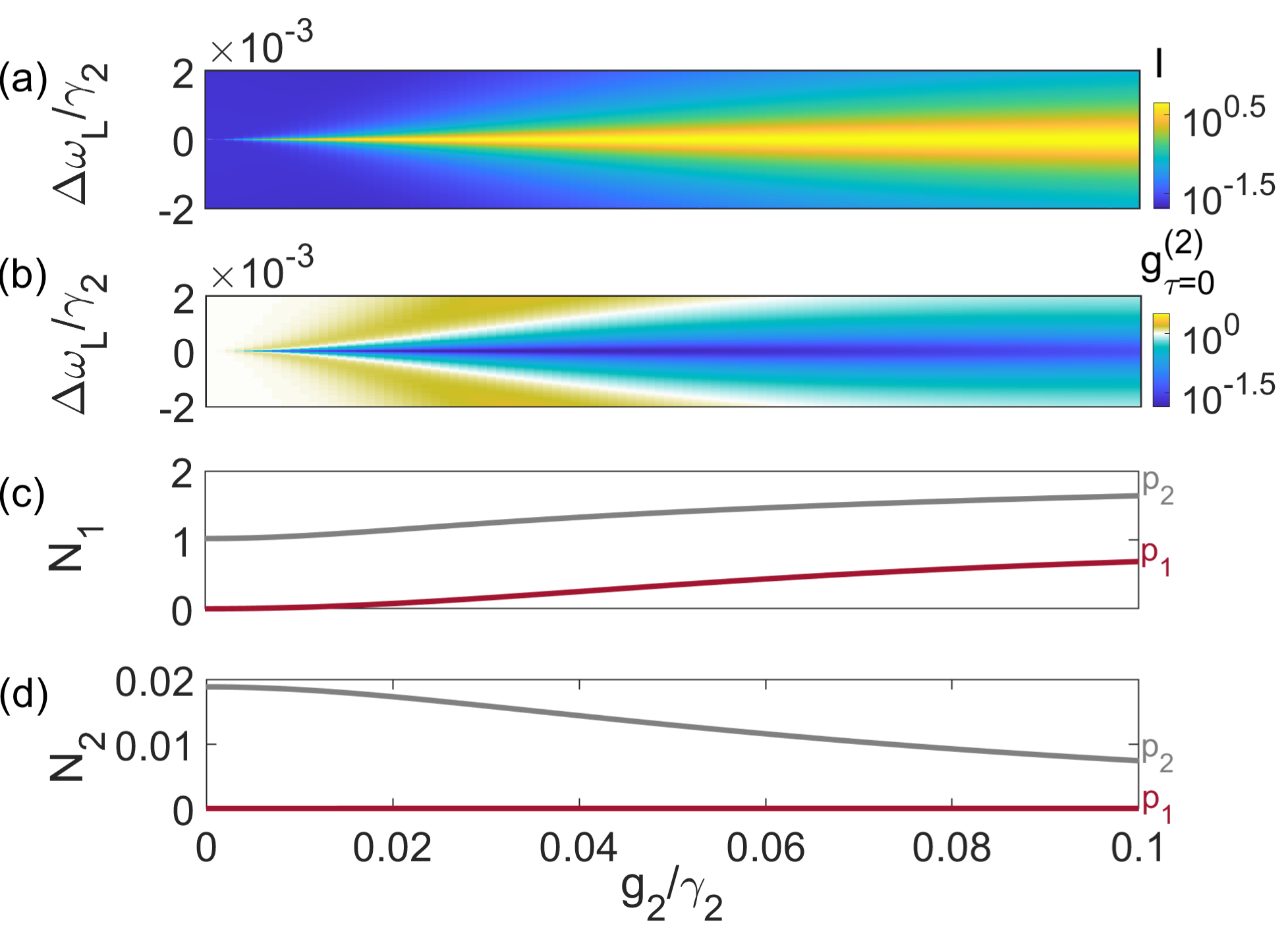}
\caption{(a)  $I$ and (b) $g_{\tau=0}^{(2)}$  as a function of   $\Delta\omega_L=\omega_e-\omega_L$ and   $g_2$. (c)  $N_1=|(j| a_1^\dagger a_1|j)|$ and (d)  $N_2=|(j| a_2^\dagger a_2|j)|$, in $|p_1)$ ($j=p_1$, dark red) and in $|p_2)$ ($j=p_2$, grey)  as a function of $g_2$.
}\label{fig:S1}
\end{figure}

\begin{figure}[ht]
\centering
\includegraphics[width=0.5\textwidth]{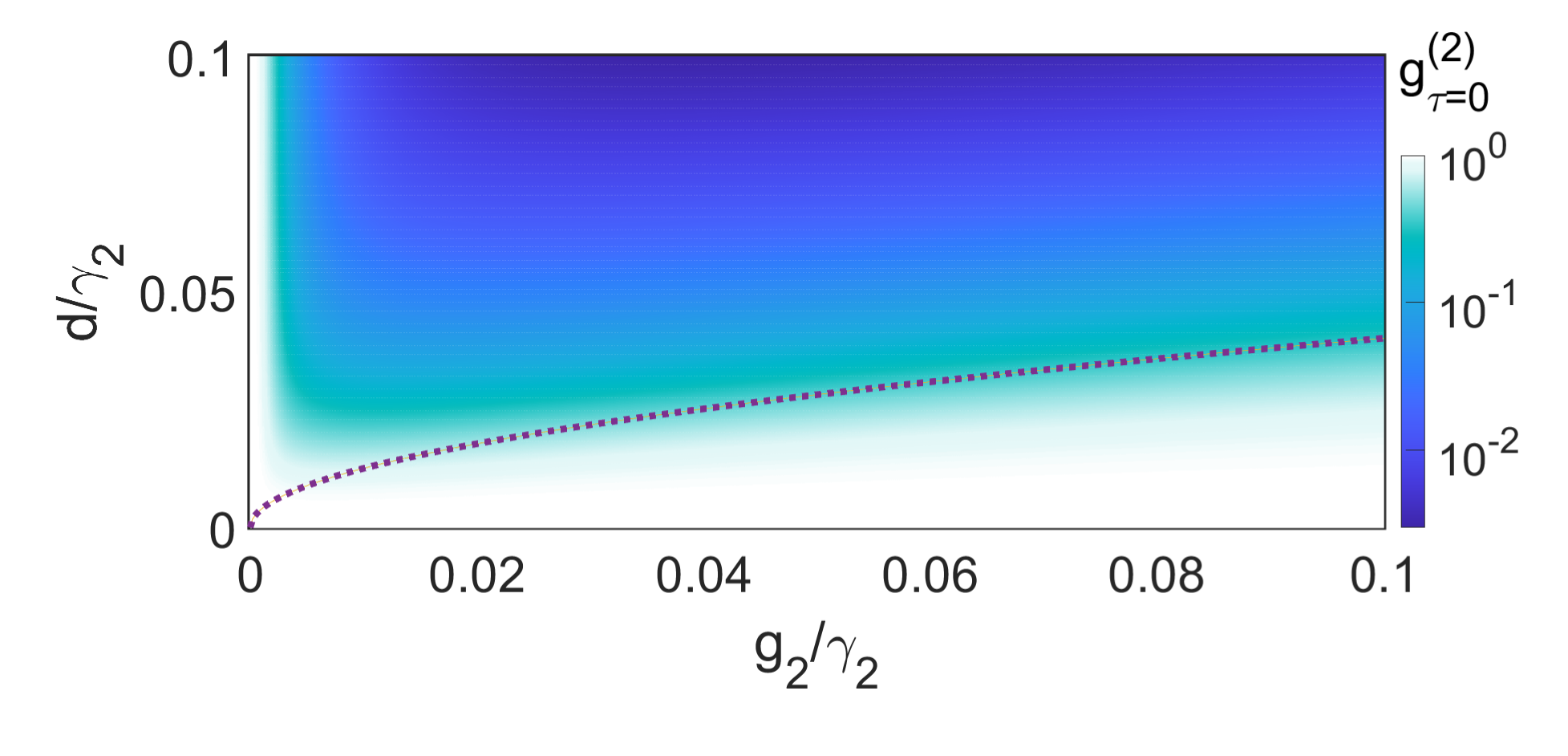}
\caption{$g_{\tau=0}^{(2)}$   as a function of $d$ and   $g_2$. The dotted line corresponds to $d=(\frac{\gamma_1}{\gamma_2}(g_2^4+\frac{g_2^2\gamma_2^2}{4}))^\frac{1}{4}$.
}\label{fig:S2}
\end{figure}

\autoref{fig:S1} studies the effect of $g_2$ on the antibunching presented in Fig.~3 and Fig.~4 in the main text, when $d=\frac{\gamma_2}{15}$ (and the other parameters are as in Fig.~3 in the main text). 
\autoref{fig:S1}(a) and \autoref{fig:S1}(b) present the intensity $I$ and the normalized zero-delayed second-order correlation function $g_{\tau=0}^{(2)}$. Moreover, to present the effect of $g_2$ on the components in $|p_1)$ and $|p_2)$, \autoref{fig:S1}(c) and \autoref{fig:S1}(d) depicts the components of the Fabry-Pérot and the plasmonic modes, $N_1$ and $N_2$, respectively, in  $|p_1)$ and $|p_2)$ as a function of $g_2$. 
In agreement with  the above-mentioned zero-order eigenvector of $|p_1)$, the dark red line in \autoref{fig:S1}(c) shows that when $g_2$ is relatively small, the component of the Fabry-Pérot mode  in $|p_1)$ is negligible. As a result, since \autoref{fig:S1}(a) and \autoref{fig:S1}(b) presents the photon statistics of the emission from the Fabry-Pérot mode, the detection of the single-photon emission is prevented.
Then, there is a range of values of $g_2$ for which the antibunching is robust.
However, when $g_2$ becomes too large,  the antibunching is suppressed (\autoref{fig:S1}(b)). 
This suppression occurs because increasing $g_2$, while keeping $d$  constant, decouples the state $a_1^\dagger a_1^\dagger|0\rangle$ from the states that involve excitation of the plasmonic mode, as is shown by the grey line in \autoref{fig:S1}(d). Consequently, $|p_2)$ is narrow has small decay rate, preventing the NHPB mechanism.
\autoref{fig:S2} investigates the interplay between $g_2$ and $d$ and presents  $g_{\tau=0}^{(2)}$ as a function of  $g_2$ and $d$ when $\omega_L=\omega_e$ (and the other parameters are as in \autoref{fig:S1}). The minimal values of $d$ that can be used to induce antibunching through the NHPB mechanism  are depicted  in \autoref{fig:S2} as a function of $g_2$ by the dotted line, derived when equating Eq.~(5) in the main text  to $1$ when $\frac{d}{g_2}\to0$.

\subsection{Robustness with respect to $g_1$}
\begin{figure}[ht]
\centering
\includegraphics[width=0.5\textwidth]{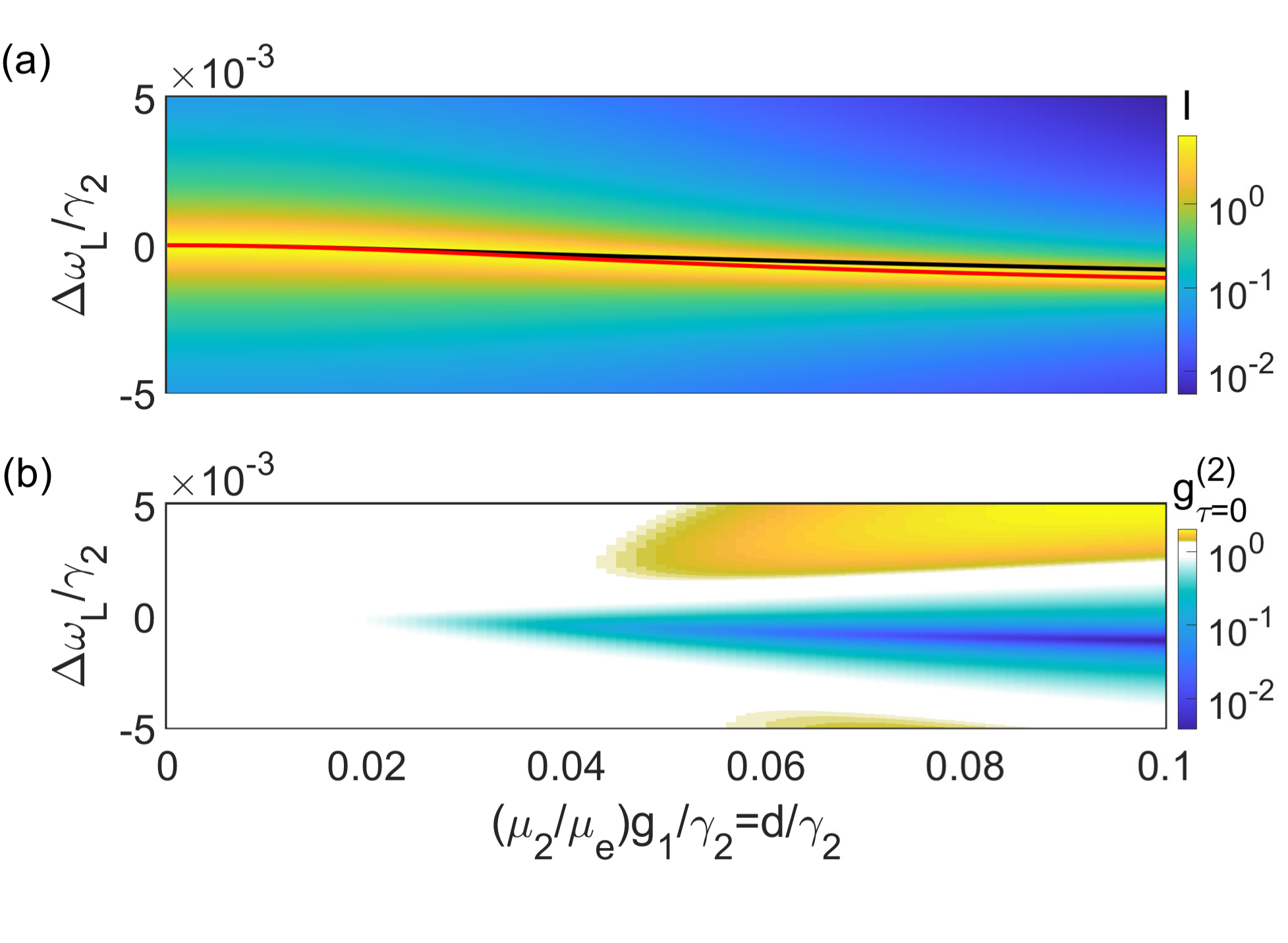}
\caption{(a)  $I$ and (b) $g_{\tau=0}^{(2)}$   as a function of   $\Delta\omega_L=\omega_e-\omega_L$ and  d, which is proportional to  $g_1$. The used ratio between $d$ and $g_1$ is $\frac{\mu_2}{\mu_e}=83.5$.
}\label{fig:S3}
\end{figure}

Fig.~3 in  the main text presents the effect of increasing $d$ on $I$ and $g_{\tau=0}^{(2)}$, while $g_1$ is kept constant. 
However, when mode 1 and mode 2 are associated with the  plasmonic and the  Fabry-Pérot modes, respectively, $g_1$ is in principle proportional to $d$: while $g_1$ is given by the interaction of the Fabry-Pérot electric field with the TLE dipole moment $\mu_e$, $d$ is given by the interaction of the Fabry-Pérot electric field with the plasmonic dipole moment $\mu_2$.
Consequently, when increasing the Fabry-Pérot electric field, both $d$ and $g_1$ are increased with the fixed ratio $\frac{d}{g_1}=\frac{\mu_2}{\mu_e}$. 
\autoref{fig:S3} presents $I$ and $g_{\tau=0}^{(2)}$ for this physical case, with all other parameters set to the same values as in Fig.~3 in the main text. 
Since $g_1$ is still smaller than $d$ and the conditions in \autoref{eq:NHPB_conds} are fulfilled, \autoref{fig:S3} shows strong antibunching. 
The black and the red lines in \autoref{fig:S3}(a) show, for each $d$, the pumping detuning that is required to populate $|p_1)$ and $|p_2)$, i.e., $\frac{E_{p_1}-\omega_L}{\gamma_2}$ and $\frac{\frac{E_{p_2}}{2}-\omega_L}{\gamma_2}$, where $E_{p_1}$ and $E_{p_2}$ are the real part of their eigenenergies.
They show that Rabi splitting and  emission in non-zero pumping detuning occur since $g_1$ reaches the strong-coupling regime for the TLE and the Fabry-Pérot mode. 

\subsection{Robustness with respect to  detuning  the plasmonic mode}

\begin{figure}[ht]
\centering
\includegraphics[width=0.5\textwidth]{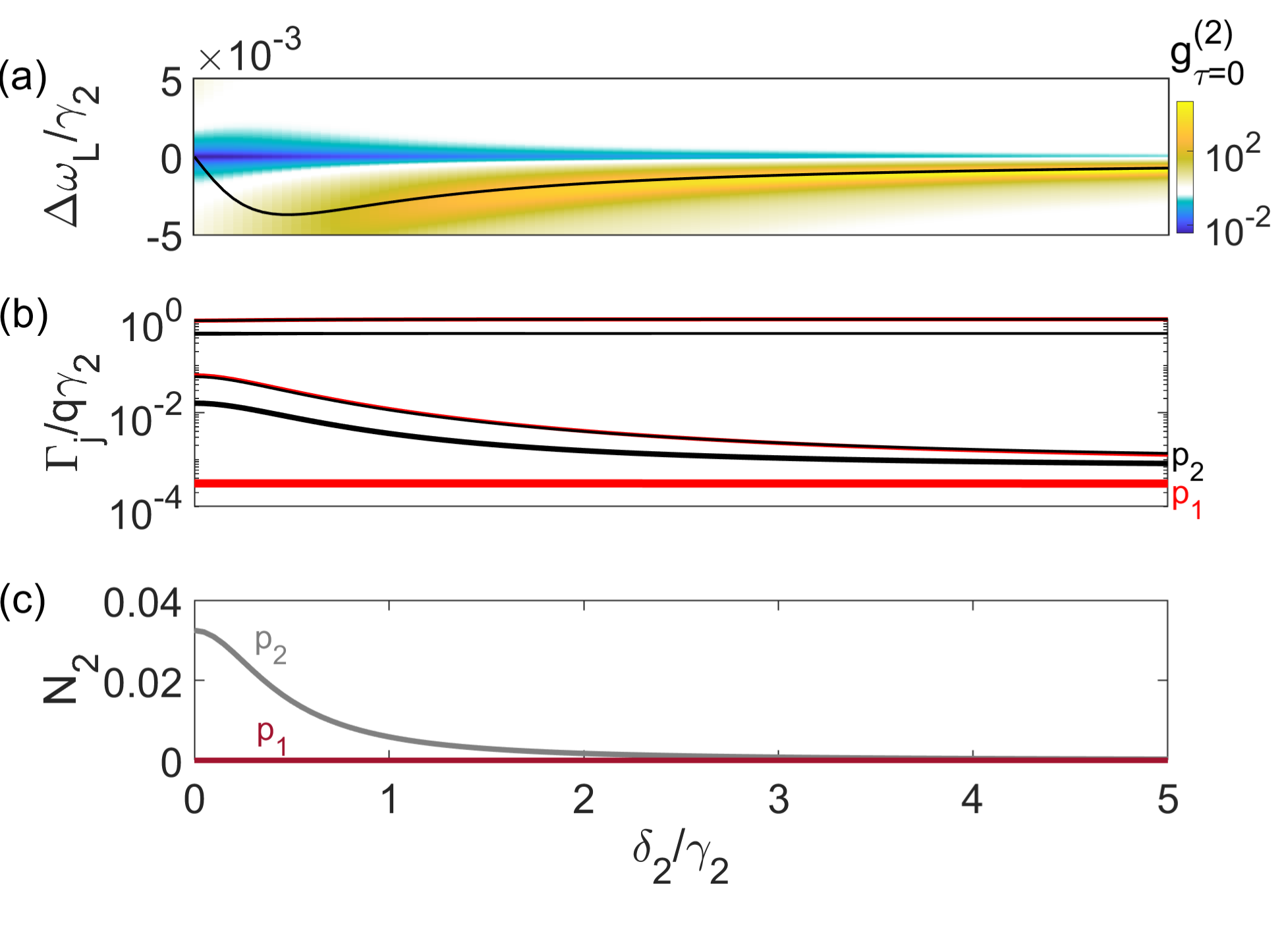}
\caption{(a)  $g_{\tau=0}^{(2)}$  as a function of  $\Delta\omega_L=\omega_e-\omega_L$ and   $\delta_2=\omega_2-\omega$. (b) $\Gamma_{j}/q$ of the first- (red, $q=1$) and the second- (black, $q=2$) excitation eigenstates as a function of $\delta_2$. (c)  $N_2=|(j| a_2^\dagger a_2|j)|$ in $|p_1)$ ($j=p_1$, dark red) and in $|p_2)$ ($j=p_2$, grey) as a function of $\delta_2$.
}\label{fig:S4}
\end{figure}

\autoref{fig:S4} studies the robustness of  the antibunching presented in  Fig.~3 and Fig.~4 in the main text when detuning the plasmonic mode from the TLE and Fabry-Pérot mode by $\delta_2$, i.e., when $\omega_2=\omega_e+\delta_2$ in Eq.~(4) in the main text. The other parameters are as in Fig.~4 in the main text.
\autoref{fig:S4}(a) shows  $g_{\tau=0}^{(2)}$; \autoref{fig:S4}(b) shows the decay rates of the first- (red) and second- (black) excitation eigenstates, where those of $|p_1)$ and $|p_2)$ are in thicker lines; and \autoref{fig:S4}(c) shows the component of the plasmonic mode $N_2$ in $|p_1)$ and $|p_2)$. 
As can be seen from \autoref{fig:S4}(a), the antibunching is robust for values of $\delta_2$ that are smaller or comparable to the plasmonic decay rate $\gamma_2$.
Since the plasmonic mode is decoupled from $|p_1)$ as is demonstrated by the dark red line in \autoref{fig:S4}(c), its detuning affects only $|p_2)$. 
The black line in \autoref{fig:S4}(a) depicts for each value of $\delta_2$ the  pumping detuning that is required  to populate  $|p_2)$, i.e., $\frac{\frac{\tilde{E}_{p2}}{2}-\omega_e}{\gamma_2}$, and shows that the detuning shifts the energy of $|p_2)$. Consequently, the anharmonicity between $|p_1)$ and $|p_2)$ is not only in the imaginary part of their eigenenergies,  shown in \autoref{fig:S4}(b), but also in their real part. Therefore, the  antibunching  in \autoref{fig:S4}(a) stems from  both the Hermitian PB and the NHPB mechanisms. 
However, the anharmonicity is destroyed and the antibunching is suppressed for large values of  $\delta_2$.
The grey line in \autoref{fig:S4}(c) shows that when $\delta_2$ becomes too large, the plasmonic mode is decoupled not only from $|p_1)$ but also from $|p_2)$, forming a narrow $|p_2)$ (thick black line in \autoref{fig:S4}(b)) that prevents the occurrence of the NHPB mechanism.

\subsection{Robustness with respect to  detuning  the TLE}

\begin{figure}[ht]
\centering
\includegraphics[width=0.5\textwidth]{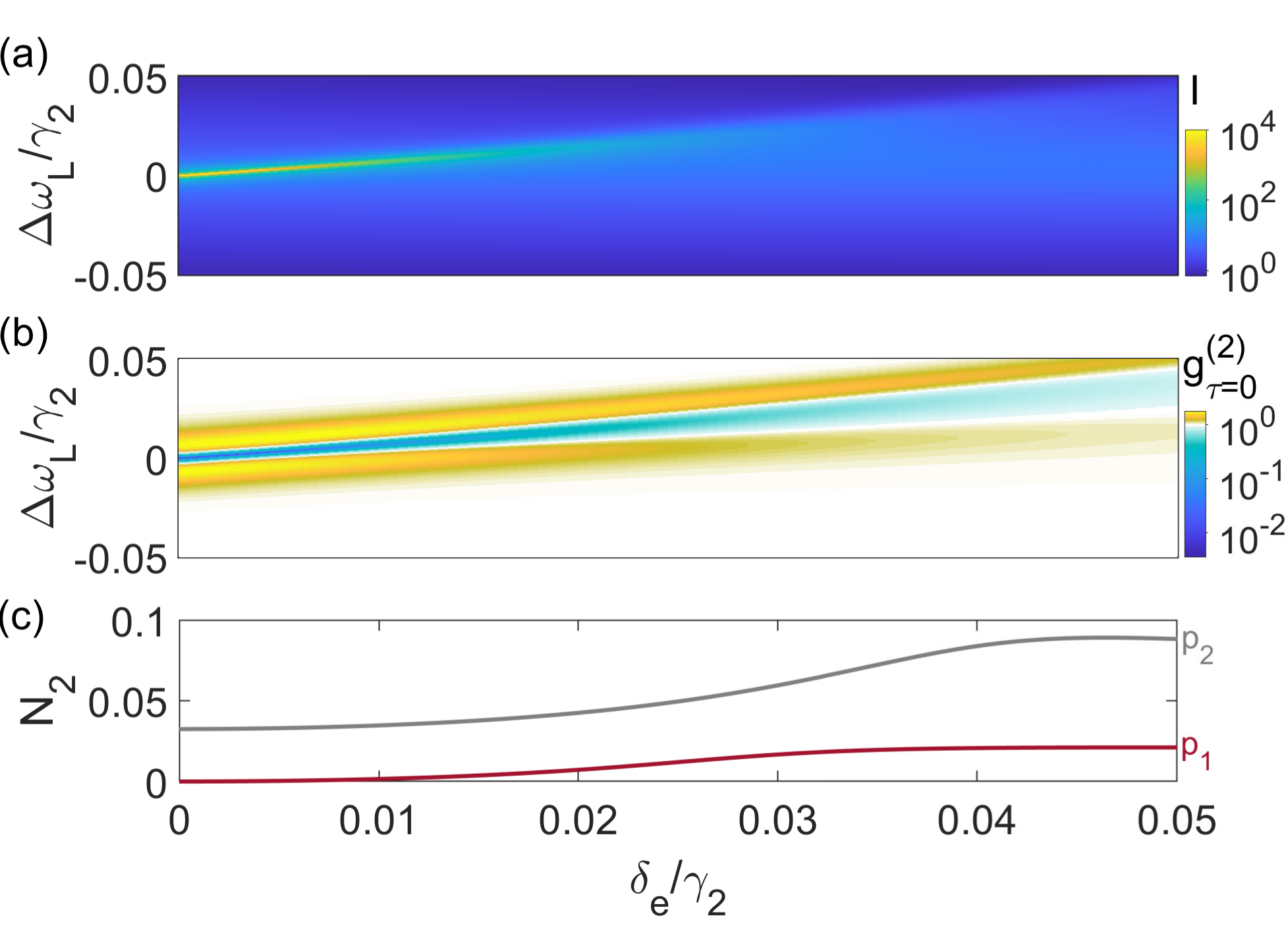}
\caption{ (a)  $I$ and (b)  $g_{\tau=0}^{(2)}$   as a function of  $\Delta\omega_L=\omega_1-\omega_L$ and  $\delta_e=\omega_e-\omega_1$. (c)  $N_2=|(j| a_2^\dagger a_2|j)|$ in $|p_1)$ ($j=p_1$, dark red) and in $|p_2)$ ($j=p_2$, grey) as a function of  $\delta_e$.
}\label{fig:S5}
\end{figure}

\autoref{fig:S5} studies the robustness of  the antibunching presented in  Fig.~3 and Fig.~4 in the main text when detuning the TLE from the optical modes by $\delta_e$, i.e., when $\omega_e=\omega_1+\delta_e$ ($\omega_1=\omega_2$) in Eq.~(4) in the main text. The other parameters are as in Fig.~4 in the main text.
\autoref{fig:S5}(a) shows $I$; \autoref{fig:S5}(b) shows $g_{\tau=0}^{(2)}$; and \autoref{fig:S5}(c) shows the component of the plasmonic mode $N_2$ in $|p_1)$ and $|p_2)$. 
As can be seen from \autoref{fig:S5}(b), the antibunching is robust when $\delta_e$ is relatively small but it is suppressed for larger values. 
Contrary to detuning the plasmonic mode (shown in \autoref{fig:S4}), which is decoupled from $|p_1)$, detuning the TLE affects both $|p_1)$ and $|p_2)$. Therefore, increasing $\delta_e$ shifts the pumping detuning that yields the maximal intensity (\autoref{fig:S5}(a)) and the antibunching (\autoref{fig:S5}(b)). Importantly, the dark red line in \autoref{fig:S5}(c) shows that relatively large $\delta_e$ (that is larger by more than an order of magnitude than the Fabry-Pérot decay rate $\gamma_1$) destroys the decoupling of the plasmonic mode from $|p_1)$, thus, forming a relatively broad $|p_1)$. As a result, the NHPB mechanism and the antibunching are suppressed.

%